\documentclass[aps,prl,reprint,groupedaddress]{revtex4-2}
\pdfoutput=1
\usepackage{dcolumn}
\usepackage{bm}
\usepackage[utf8]{inputenc}
\usepackage[T1]{fontenc}
\usepackage{booktabs, array, mathptmx, float, tabularx, booktabs, lipsum, amsmath,multirow,amssymb}
\usepackage{siunitx, xcolor}
\usepackage[version=4]{mhchem}
\usepackage{graphicx,amssymb}
\usepackage{subfigure}
\usepackage{indentfirst,latexsym,bm}
\usepackage{threeparttable}
\graphicspath{{figs/}{figsgaoerb/}}
\usepackage[colorlinks,linkcolor=blue,anchorcolor=blue,citecolor=blue]{hyperref}
\usepackage{hyperref}
\usepackage{amsfonts}
\usepackage{newtxtext,newtxmath}
\usepackage{appendix}
\begin{document}

\title{Stable Long-Range Charging and Boundary-Mediated Thresholds in a Waveguide Quantum Battery}
\author{Li-Hong Yu$^{1}$}
\author{Zhuo-Yu Zhang$^{1}$}
\author{Yu Wang$^{1}$}
\author{Xin Li$^{1}$}
\email{To whom correspondence should be addressed. Email: lixin$_$physics@126.com}
\affiliation{$^{1}$Data science research center, Kunming University of Science and Technology, Kunming, 650093, China}


\begin{abstract}
Boundary effects and bound states in structured reservoirs are ubiquitous features of open quantum systems. We investigate a minimal charging setup consisting of two emitters coupled to a semi-infinite waveguide and identify the optimal charging mechanism, in which non-Markovian 
dynamics driven by out-of-band bound states give rise to dissipationless, single-frequency oscillations of the stored energy. We show that near-unity charging efficiency  at arbitrary distances is achievable under weak to moderate coupling via detuning control, provided that the battery-to-boundary separation $n$ is less than three times the charger-to-boundary separation $m$ ($n<3m$)---a geometric constraint imposed by mirror-induced feedback---while the average charging power decays exponentially with the distance $n-m$. These findings can be directly tested on superconducting circuit platforms, with implications for energy transfer and photon routing in open waveguide architectures.
\end{abstract}
\maketitle
\section{INTRODUCTION}
Recent advances in quantum technologies---including quantum computing, sensing, simulation, and communication---are driving transformative global development \cite{turyshev2026quantumtechnologiessystemlevelperformance, Wang2025, 10828532,
	doi:10.1126/science.adz8659,10.1093/nsr/nwaf289}. In this context, quantum batteries (QBs) have been proposed and have since bridged the gap between quantum thermodynamics and quantum information science \cite{PhysRevE.87.042123Alicki}. By leveraging entanglement and coherence, QBs can achieve superlinear scaling of charging power in many-body systems beyond classical limits \cite{PhysRevE.87.042123Alicki,PhysRevE.97.062105,doi:10.1126/sciadv.abk3160,RevModPhys.96.031001,PhysRevResearch.4.033216,Campaioli2018,Binder_2015,Ferraro2026,Zahia_2025,farimani2026detuningcontrolledphasetransitionpassive, PhysRevLett.120.117702,PhysRevB.99.205437}, a hallmark of the superabsorption effect that was theoretically proposed \cite{Higgins2014} and  
experimentally verified \cite{doi:10.1126/sciadv.abk3160,yang2021realization}.
Despite these prospects, practical implementation remains hindered by low capacity, short retention times, and environmental decoherence. Addressing these bottlenecks requires recognizing that macroscopic charging dynamics are governed by the collective quantum behavior of two-level systems (TLSs) \cite{PhysRevLett.118.150601,595w-rwr4}. As TLSs serve as the elementary building blocks of more complex QBs, understanding energy transfer in minimal TLS-based QB configurations is a prerequisite for scalable energy storage.

Solid-state architectures are widely recognized as the essential path toward the ultimate realization of portable QBs. Current mainstream solid-state platforms include semiconductor quantum dots, diamond nitrogen-vacancy (NV) centers, and superconducting quantum circuits (SQCs).  Coupling between double quantum dots has been exploited to explore charging dynamics and correlation effects in energy storage \cite{https://doi.org/10.1002/qute.70375,loukhssami2026quantum}, while NV centers have been utilized both to construct QBs
with self-discharge suppression characteristics and to enhance coherent ergotropy via hyperfine-controlled quantum coherence \cite{d9k1-75d4}. SQCs offer distinct advantages over the former two.  First, the transition frequencies of their artificial TLSs (e.g., transmons) and the coupling between them can be tuned more flexibly via external magnetic flux or microwave bias \cite{PhysRevA.76.042319,Elghaayda2025}, exhibiting higher control fidelity and nanosecond-level response times \cite{Barends2014,RevModPhys.93.025005}. Second, Josephson junctions (JJs) are fully compatible with mature micro- and nanofabrication techniques, allowing chip-scale integration \cite{kim2026advancesjosephsonjunctionmaterials,Tolpygo_2020,10798992}.  Current research on JJ-based QBs primarily employs short-range energy exchange mechanisms between a quantum charger (QC) and the QB, including on-chip mutual inductive coupling  \cite{varrica2026quantumbatteriestwodimensionalmaterialbased}, capacitive coupling \cite{y3qx-cs3r}, and nearest-neighbor direct interactions \cite{sp5l-c6m8}. However, the spatial separation between the QC and the QB is crucial for enhancing the flexibility and scalability of  integrated circuit layouts. In addition,  contactless energy transfer mitigates heat leakage \cite{raicu2025cryogenic,tian2025high} and control-line noise \cite{yx15-jyl7} in cryogenic systems, while waveguide-confined states suppress radiative dissipation into continuous-spectrum channels.

Although remote charging of qubit batteries has been explored in coupled-cavity arrays \cite{7c9x-rhld}, hollow rectangular
waveguides \cite{PhysRevLett.132.090401}, and topological photonic structures \cite{PhysRevLett.134.180401}, whether high-efficiency energy transmission can be maintained at arbitrary distances under more practical constraints remains elusive. Moreover, most schemes still assume an infinite or periodic waveguide, which simplifies the analysis via translational symmetry. In experiments, however, the waveguide terminates at a physical boundary (e.g., reflector, cutoff) \cite{PhysRevLett.122.073601,PhysRevResearch.6.043121,PhysRevApplied.20.024058}, introducing a new degree of freedom: the emitter–boundary distance. This geometry couples incident, reflected, and multiply reflected waves within the same channel, and imparts  a round-trip delay and structured band edges. Such configurations give rise to effects absent in infinite waveguides, such as boundary-mediated suppression or trapping of spontaneous emission \cite{PhysRevA.87.013820}, atom–photon bound states \cite{PhysRevA.87.013820, PhysRevA.109.023712}, reflected photon bunching \cite{PhysRevA.91.053845}, and non-Markovian dynamics arising from finite bandwidth and feedback-induced memory \cite{Zeng_2023}. Therefore, the role of boundaries in energy transfer warrants special attention.

 In this paper, we consider a 1D semi-infinite coupled-resonator waveguide (CRW) as a quantum bus, which is terminated by a perfect mirror at the end proximal to the emitter.
Its cosine-type dispersion relation is characteristic of JJ resonator arrays under the tight-binding approximation \cite{PhysRevX.12.031036}. By tracing the origins of out-of-band bound states, we delineate the parameter regime for stable, long-distance charging. The scaling laws established here provide practical guidance for the design of dynamic energy routing and distributed quantum energy-transfer networks, provided that the distal end of a finite-length JJ resonator array does not appreciably perturb the spatial profiles of the relevant bound states. Conversely, a systematic analysis reveals that when additional out-of-band bound states emerge with comparable weights, the resulting multimode resonances render the determination of the optimal charging time highly nontrivial, while any coexisting in-band bound states cannot support robust energy transfer.

\section{Charging Protocol}
\begin{figure}
	\centering
	\includegraphics[width=1\linewidth]{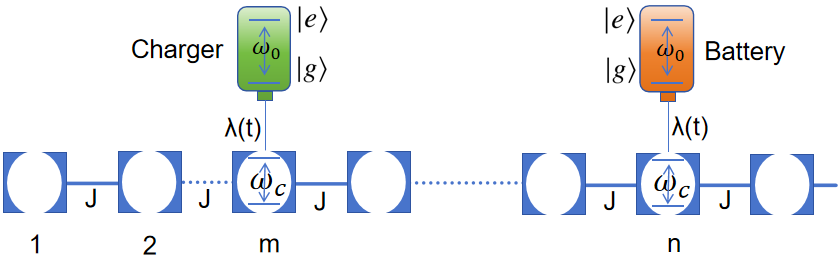}
	\caption{Schematic of the charging scheme. The QC and QB are coupled through a 1D semi-infinite CRW, where 
	$m$ and $n$ refer to their respective positions. The time-dependent interaction strength is given by $\lambda(t)=\lambda \Theta(\tau-t)$, with the step function 
	$\Theta$ satisfying
	$\Theta=1$ for $t\in[0,\tau]$ and  $\Theta=0$ otherwise.}
	\label{Fig1}
\end{figure}
Our charging scheme is mediated by a semi-infinite  CRW comprising $N$ resonators with open boundary conditions (OBCs), which serves as the energy transfer channel between the QC and the QB, as illustrated in Fig.~\ref{Fig1}.  The total Hamiltonian of the system reads (set $\hslash=1$):
\begin{equation}\label{1}
	\begin{aligned}
		H&=H_{C}+H_{B}+H_{CW}+H_{BW}+H_{W},\,\,with
		\\ H_{x}&=\omega_{0}\sigma^{\dag}_{x}\sigma_{x}\,\,\,(x=C\,\, or\,\, B),
		\\H_{CW}&=\lambda(t)(\sigma^{\dag}_{C}A_{m}+h.c.) 
		,H_{BW}=\lambda(t)(\sigma^{\dag}_{B}A_{n}+h.c.),\,and	\\H_{W}&=\sum_{j=1}^{N}\omega_{c}A^{\dag}_{j}A_{j}+\sum_{j=1}^{N-1}J(A^{\dag}_{j}A_{j+1}+h.c.).\,\,\,\,\,\,\,\,\,\,\,\,\,\,\,\,\,\,\,\,\,\,\,\,\,\,\,\,\,\,\,\,\,\,\,\,\,\,\,\,\,\,\,\,\,\,\,\,\,\,\,\,\,\,\,\,\,\,\,\,
	\end{aligned}
\end{equation}
Here, $\omega_{c}$ is the resonator frequency, and $J$ is the nearest-neighbor hopping amplitude between adjacent resonators. For simplicity, we focus on the case where the QC at site $m$ and the QB at site $n$ share an identical resonance frequency $\omega_{0}$ and are coupled to the CRW with a common Jaynes–Cummings interaction strength 
$\lambda(t)$. Numerical simulations confirm that assigning distinct resonance frequencies or interaction strengths to the QC and QB offers no additional advantage for the charging performance. The Heaviside step function
 $\Theta(\tau-t)$ in the definition $\lambda(t)\equiv\lambda \Theta(\tau-t)$ serves to abruptly turn off the interaction at $t=\tau$, the instant at which the QB reaches its maximum energy. 

The commutation relation $[H,\Sigma_{j=1}^{N}A_{j}^{\dag}A_{j}+\sigma_{C}^{\dag}\sigma_{C}+\sigma_{B}^{\dag}\sigma_{B}]=0$ guarantees the conservation of the total excitation number under unitary evolution. Applying the Fourier transform $A_{j}=\Sigma_{k=1}^{N}{\sqrt\frac{2}{N+1}}A_{k}\sin (\frac{jk\pi}{N+1})$,
the interaction terms become	
$H_{CW}+H_{BW}=\Sigma_{k=1}^{N}{\sqrt\frac{2\lambda^{2}}{N+1}}[\sigma^{\dag}_{C}A_{k}\sin (\frac{mk\pi}{N+1})+\sigma^{\dag}_{B}A_{k}\sin (\frac{nk\pi}{N+1})+h.c.]$, while
the CRW Hamiltonian reduces to the diagonal form $H_{W}=\Sigma_{k}\varOmega_{k}A^{\dag}_{k}A_{k}$, with the dispersion  $\varOmega_{k}=\omega_{c}+\omega_{k}=\omega_{c}+2J\cos(\frac{\pi k}{N+1})$ \cite{PhysRevA.78.042304}. It is assumed that at $t=0$, the QC is maximally excited, the QB is in its ground state, and the CRW is in the vacuum state, so that the global initial state reads $|\varphi_{0}\rangle=|e_{C}\rangle|g_{B}\rangle|vac\rangle$, where $|e\rangle$ ($|g\rangle$) denotes the excited (ground) states of the QC or QB, and $|vac\rangle$ is the CRW vacuum. Within the single-excitation subspace,  the exact time-evolved state takes the form
\begin{equation}\label{2}
	\begin{aligned}
		|\varphi(t)\rangle=e^{-i\omega_{c}t}[C(t)|e_{C}\rangle |g_{B}\rangle |vac\rangle+\,\,\,\,\,\,\,\,\,\,\,\,\,\,\,\,\,\,\,\,\,\,\,\,\,\,\,\,\,\,\,\,\,\,\,\,\,\,\,\,\,\,\,\,\,\,\,\,\,\,\,\,\,\,\,\,\,\,\,\,\,\,\,\,\,\\
		B(t)|g_{C}\rangle|e_{B}\rangle|vac\rangle	+\sum_{k}a_{k}(t)e^{-i\omega_{k}t}|g_{C}\rangle |g_{B}\rangle|1^{A}_{k}\rangle],\,\,\,\,\,\,\,\,\,
	\end{aligned}
\end{equation}
with $a_{k}(t)$ representing
the amplitudes of propagating mode $|1_{k}^{A}\rangle(\equiv A_{k}^{\dagger}|vac\rangle)$.

To quantify the energy transferred from the QC to the QB, we evaluate the instantaneous stored energy of the QB,
\begin{equation}\label{3}
	E_{B}(t) =Tr\left[ \rho_B(t) H_B \right]\equiv\omega_0|B(t)|^{2},
\end{equation}
where $\rho_B(t)=Tr_{C+W}[|\varphi(t)\rangle\langle\varphi(t)|]$ is the reduced density matrix of the QB, obtained by tracing out the QC and the CRW. The first local maximum of $E_{B}(t)$, occurring at $t=\tau_{1}$, constitutes the most operationally relevant figure of merit for the charging process \cite{PhysRevLett.120.117702,PhysRevB.98.205423, PhysRevResearch.4.033216}. Another key performance metric is the average charging power
\begin{equation}\label{5}
	\overline{\mathcal{P}} =\frac{1}{\tau_{1}}\int_{0}^{\tau_{1}}\frac{d}{dt}Tr[\rho_B(t) H_B] =\frac{E_B(\tau_{1})}{\tau_{1}},
\end{equation}
which quantifies the overall speed of the energy transfer. Finally, since not all energy stored in the QB can be extracted as useful work, the ergotropy must be strictly positive
 \cite{PhysRevLett.125.180603,PhysRevE.102.042111,PhysRevLett.131.030402}: 
\begin{equation}\label{4}
	 \mathcal{E}(\tau_{1})=\omega_0 ( 2 |B(\tau_{1})|^2 - 1 )\Theta(|B(\tau_{1})|^2 -\frac{1}{2}) >0,
\end{equation}
which imposes a nontrivial physical constraint on $B(\tau_{1})$.

\section{~~FORMAL SOLUTION}
With Eq.~(\ref{2}),  the Schrödinger equation $H|\varphi(t)\rangle=i\partial_{t}|\varphi(t)\rangle$ yields the following coupled equations for the probability amplitudes:
\begin{subequations}\label{6}
	\begin{align}
	i\dot{C}(t)=\omega C(t)+\Sigma_{k}{\sqrt\frac{2\lambda^{2}}{N+1}}\sin (\frac{mk\pi}{N+1})e^{-i\omega_{k}t}a_{k}(t),\label{6a}\,\,\,\,\,\,\,\,\,\,\,\,\,\,\,\,\,\,\,\,\,\,\,\,\,\,\\ 
	i\dot{B}(t)=\omega B(t)+\Sigma_{k}{\sqrt\frac{2\lambda^{2}}{N+1}}\sin (\frac{nk\pi}{N+1})e^{-i\omega_{k}t}a_{k}(t),\label{6b}\,\,\,\,\,\,\,\,\,\,\,\,\,\,\,\,\,\,\,\,\,\,\,\,\,\,\\ 
	i\dot{a}_{k}(t)={\sqrt\frac{2\lambda^{2}}{N+1}}e^{i\omega_{k}t}[\sin (\frac{mk\pi}{N+1})C(t)+\sin (\frac{nk\pi}{N+1})B(t)]\label{6c}.
	\end{align}
\end{subequations}
Setting $J$ as the energy unit,  Eq.~(\ref{6}) shows that the validity of this charging protocol depends  on the atom-photon detuning $\omega=\omega_{0}-\omega_{c}$,  the interaction strength $\lambda$, and the positions of the QC and QB. Substituting the integral form of Eq.~(\ref{6c}),
\begin{equation}\label{7}
	\begin{aligned}
		&a_{k}(t)=\\
		&-i{\sqrt\frac{2\lambda^{2}}{N+1}}\int_{0}^{t}e^{i\omega_{k}t'}[\sin (\frac{mk\pi}{N+1})C(t')+\sin (\frac{nk\pi}{N+1})B(t')]dt',\\
	\end{aligned}
\end{equation}
into Eq.~(\ref{6a}) and Eq.~(\ref{6b}),
we obtain two coupled integro-differential equations for $C(t)$ and $B(t)$:
\begin{widetext}\label{8}
	\begin{subequations}
		\begin{align}
\dot{C}(t)=-i\omega C(t)-{\frac{2\lambda^{2}}{N+1}}\Sigma_{k}\int_{0}^{t}[\sin^{2} (\frac{mk\pi}{N+1})C(t')+\sin (\frac{nk\pi}{N+1})\sin (\frac{mk\pi}{N+1})B(t')]e^{i\omega_{k}(t'-t)}dt',\label{8a}
\\ \dot{B}(t)=-i\omega B(t)-{\frac{2\lambda^{2}}{N+1}}\Sigma_{k}\int_{0}^{t}[\sin (\frac{nk\pi}{N+1})\sin (\frac{mk\pi}{N+1})C(t')+\sin^{2} (\frac{nk\pi}{N+1})B(t')]e^{i\omega_{k}(t'-t)}dt'.\label{8b}
\end{align}
 \end{subequations}
  \end{widetext}
Taking the Laplace transformation of Eq.~(\ref{8a}) and Eq.~(\ref{8b}) yields the following algebraic system in the Laplace domain:
\begin{subequations}\label{9}
	\begin{align}
		C(s)s-1=-i\omega C(s)-\lambda^{2}[Q(m,m,s)C(s)+Q(m,n,s)B(s)],\label{9a}\\
		B(s)s=-i\omega B(s)-\lambda^{2}[Q(m,n,s)C(s)+Q(n,n,s)B(s)],\label{9b}
	\end{align}
\end{subequations}
where the kernel function in the large-$N$ limit is explicitly defined as 
\begin{equation}\label{10}
Q(m,n,s)=\frac{1}{\pi}\int^{\pi}_{-\pi}\frac{\sin (mk)\sin (nk)}{s+2iJ\cos(k)}dk.
\end{equation}	
 Solving the linear system of Eq.~\eqref{9} gives the closed-form solutions for $C(s)$ and $B(s)$:
\begin{widetext}
\begin{subequations}\label{11}
	\begin{align}
		C(s)=\frac{-(\lambda^{2}Q(n,n,s)+s+i\omega)}{\lambda^{4}Q(m,n,s)^{2}-(\lambda^{2}Q(n,n,s)+s+i\omega)(\lambda^{2}Q(m,m,s)+s+i\omega)},\label{11a}\\
		B(s)=\frac{\lambda^{2}Q(m,n,s)}{\lambda^{4}Q(m,n,s)^{2}-(\lambda^{2}Q(n,n,s)+s+i\omega)(\lambda^{2}Q(m,m,s)+s+i\omega)}.\label{11b}
	\end{align}
\end{subequations}	
\end{widetext}
Note that the kernel $Q(m,n,s)$ admits the trigonometric decomposition
\begin{equation}\label{12}
Q(m,n,s)=F(n-m,s)-F(n+m,s),
\end{equation}
with the auxiliary function $F(a,s)=\frac{1}{2\pi}\int^{\pi}_{-\pi}\frac{e^{i k a}}{s+2 i J \cos (k)}dk$. For sufficiently large $m$, the asymptotic condition 
 $\lim\limits_{\substack{ a\to \infty }} F(a,s)=0$ holds, and Eq.~(\ref{11a}) and Eq.~(\ref{11b}) simplify to
 \begin{subequations}\label{13}
	\begin{align}
		C(s)\backsimeq\frac{1}{2}[G_{+}(s)-G_{-}(s)]\label{13a},\\
		B(s)\backsimeq\frac{1}{2}[G_{+}(s)+G_{-}(s)],\label{13b}
	\end{align}
\end{subequations}
where $G_{\pm}(s)=\frac{1}{\lambda^{2}[F(n-m,s)\pm F(0,s)]\pm(s+i\omega)}$ is precisely the propagator under periodic boundary conditions (PBCs) \cite{10.1088/1674-1056/ae3b32}.

\begin{figure}
	\centering
	\includegraphics[width=0.305\linewidth]{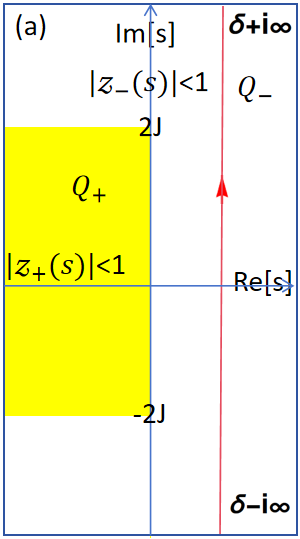}
	\includegraphics[width=0.5\linewidth]{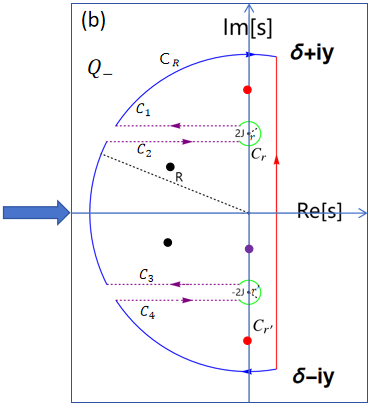}
	\caption{ Complex $s$-plane partition and contour deformation.  (a) The complex plane is divided into two analytic domains:  $Q_{+}$ (yellow, $|z_+(s)|<1$) and $Q_{-}$ (white, $|z_-(s)|<1$), as defined in Eq.~(\ref{15}).   The Bromwich contour (red line) runs from $\delta-i\infty$ to $\delta+i\infty$ and lies entirely within $Q_{-}$.  (b) Upon analytic continuation of $Q_{-}$ to the full complex plane, the integral is evaluated by contour deformation. Black dots denote unstable poles; red and purple dots indicate bound states outside and within the band, respectively.}
\label{Fig2}
\end{figure}

To retrieve the real-time dynamics, the time-domain excited-state amplitudes  $C(t)$ and $B(t)$ are recovered via the inverse Laplace transform:  
\begin{equation}\label{14}
	\begin{aligned}
	Y(t)=\frac{1}{2\pi i}\int^{\delta+i\infty}_{\delta-i\infty}Y(s)e^{st}ds,
	\end{aligned}
\end{equation}
where  $Y\in\{C,B\}$. The real constant $\delta$ is chosen such that the vertical line $s=\delta$ lies to the right of all singularities (poles and branch points) of the integrand $Y(s)$ in the complex $s$-plane, as illustrated in Fig.~\ref{Fig2}.

The variable substitution $z=e^{ik}$\ \ transforms Eq.~(\ref{10}) into a contour integral along   the unit circle on the complex $z$-plane: $Q(m,n,s)=\frac{1}{4\pi }\ointctrclockwise f(z) dz$, with$f(z)=\frac{(z^{m}-z^{-m})(z^{n}-z^{-n})}{J(z^{2}+1)-isz}$. By Vieta's formulas, the two roots of the denominator of $f(z)$, given by $z_{\pm}(s)=\frac{is}{2J}\pm i\sqrt{1+(\frac{s}{2J})^{2}}$, satisfy the identity $z_{+}(s)z_{-}(s)=1$. The residue theorem provides the following solution:
\begin{equation}\label{15} 
\begin{aligned}
	&Q(m,n,s) =\left\{
	\begin{aligned}
		\frac{i \left(z_{+}(s)^{1+m+n}-z_{+}(s)^{1+| n-m|}\right)}{J \left(z_{+}(s)^2-1\right)}\equiv Q_{+},\\
	\frac{i\left(z_{-}(s)^{1+m+n}-z_{-}(s)^{1+| n-m|}\right)}{J \left(z_{-}(s)^2-1\right)}\equiv Q_{-}.
		\end{aligned}
		\right.\\
	\end{aligned}
\end{equation}
 To perform the inverse Laplace transform for Eq.~(\ref{11a}) and Eq.~(\ref{11b}), one must properly handle the square-root singularity arising from the multivalued function $\sqrt{1+(\frac{s}{2J})^{2}}$ embedded in  $z_{\pm}(s)$. We adopt the branch-cut prescription of Ref.~\cite{PhysRevA.93.062104}, defining the cuts along the band edges  $s=x\pm2iJ$ (with $x<0$) on the complex $s$-plane parameterized by $s=x+iy$:
\begin{equation}\label{16}
	\sqrt{1+(\frac{s}{2J})^{2}}\equiv\sqrt{|1+(\frac{s}{2J})^{2}|}\exp [\frac{Arg(\frac{s}{2J}+i)+Arg(\frac{s}{2J}-i)}{-2i}].
\end{equation}
 Since the Bromwich-integral path $s\in(\delta-i\infty,\delta+i\infty)$ [the red line in Fig.~\ref{Fig2}] lies entirely within the white region,  $Q_{-}(m,n)$ can be analytically extended to the full complex plane [Fig.~\ref{Fig2} (b)]. To apply Cauchy’s theorem, we construct a closed contour that avoids the two branch cuts [Fig.~\ref{Fig2} (b)]:
\begin{widetext}
	\begin{equation}\label{17}
		\begin{aligned}
			\int^{\delta+i\infty}_{\delta-i\infty}Y(s)e^{st}ds
			=\lim\limits_{\substack{ r,r^{\prime} \to \ 0\\R\to\ \infty }}(\int_{C_{1,2,3,4}} +\int_{C_{r,r^{\prime}}}+\int_{C_{R}})Y(s)e^{st}ds+2\pi i\sum_{s_{l}}Res_{Y}[s_{l}]e^{s_{l}t},
		\end{aligned}
	\end{equation}
\end{widetext}
where $Res_{Y}[s_{l}]$ indicates that the amplitudes $C(s)$ and $B(s)$  yield distinct residues at the same pole $s_{l}$. By Jordan's lemma, the contributions from the small circular arcs $C_{r,r^{\prime}}$ and  the large outer arc $C_{R}$ vanish in the stated limit. To identify a stable charging protocol, we focus on the long-time behavior of the system. The Lebesgue-Riemann lemma guarantees that the integrals along the Hankel-type paths $C_{1,2,3,4}$ (see the Appendix for explicit expressions) decay to zero for sufficiently large $Jt$ \cite{bochner1949fourier}. Note that the integration along the $C_{2,3}$ is performed on the second Riemann sheet. Consequently, Eq.~(\ref{14}) reduces to
\begin{equation}\label{18}
 Y(Jt\gg 1)\simeq\sum_{s_{l}}Res_{Y}[s_{l}]e^{s_{l}t}.
\end{equation}
To facilitate the identification of the poles $s_{l}$, we recast Eq.~(\ref{11}) as a rational function via the substitution
 $z_{-}(s)$=$Z_{-}$ [equivalent to $s= -iJ(Z_{-}+\frac{1}{Z_{-}})$]: 
\begin{equation}\label{19}
	\begin{aligned}
		B(s)=\frac{\mathcal{D}_{b}(Z_{-})}{\mathcal{D}(Z_{-})},	C(s)=\frac{\mathcal{D}_{c}(Z_{-})}{\mathcal{D}(Z_{-})}.
	\end{aligned}
\end{equation}
 The explicit forms of the numerator and denominator polynomials are given by
\begin{widetext}\label{20}
\begin{subequations}
	\begin{align}
	\mathcal{D}_{b}(Z_{-})=&i J \lambda ^2 \left(Z_{-}^2-1\right)^2 Z_{-}^3 \left(Z_{-}^{m+n}-Z_{-}^{|n-m|}\right),\label{20a}
	\\
	\mathcal{D}_{c}(Z_{-})=&iJ\left(Z_{-}^2-1\right)^2 (J^2 Z_{-}^5-J^2 Z_{-}-J \omega  Z_{-}^4+J \omega Z_{-}^2-\lambda ^2 Z_{-}^{2 n+3}+\lambda ^2 Z_{-}^3),\label{20b}
	\\\mathcal{D}(Z_{-})=&h_0+h_1 \left(Z_{-}^{2 m}+Z_{-}^{2 n}\right)+\lambda ^4 Z_{-}^4 \left(Z_{-}^2-1\right)( 2 Z_{-}^{| n-m| +m+n}-Z_{-}^{2|n-m|}-Z_{-}^{2 m}-Z_{-}^{2 n}),\label{20c}
\end{align}
\end{subequations}
\end{widetext}
where
\begin{subequations}\label{21}
	\begin{align}
h_0(Z_{-})&=
(Z_{-}^2-1) (J^2 (Z_{-}^4-1)-J \omega  Z_{-} (Z_{-}^2-1)+\lambda ^2 Z_{-}^2)^2,\label{21a}
\\h_1(Z_{-})&=
-J \lambda ^2 Z_{-}^2 (Z_{-}-1)^2 (Z_{-}+1)^2 (J Z_{-}^2+J-\omega  Z_{-}).
\end{align}
\end{subequations}
Eq.~(\ref{19}) reveals that exchanging the positions of the QC and QB leaves the amplitude $B(t)$ invariant; as shown below, its effect on $C(t)$ is likewise negligible. We therefore adopt the convention  $m<n$ (QC placed to the left of the QB in Fig.~\ref{Fig1}) throughout. It should be noted that not all roots of $\mathcal{D}(Z_{-})=0$ correspond to physical poles $s_{l}$: only those satisfying $Z_{l}=\frac{is_{l}}{2J}- i\sqrt{1+(\frac{s_{l}}{2J})^{2}}$ are admissible. Assuming all roots are simple and applying L'Hôpital's rule, the residues are obtained via partial-fraction decomposition
\begin{equation}\label{22}
Res_{Y}[s_{l}]=J[\frac{1}{Z^{2}_{l}}-1]\frac{i\mathcal{D}_{Y}(Z_{l})}{\mathcal{D}^{\prime}(Z_{l})},
\end{equation}
where $\mathcal{D}^{\prime}(Z_{l})\equiv
\frac{\partial\mathcal{D}(Z_{-})}{\partial Z_{-}}\big|_{Z_{-}=Z_{l}}
$. Two classes of poles emerge.
The unstable poles with negative real parts govern the spontaneous emission of the QC [black dots in Fig.~\ref{Fig2}(b)]. The purely imaginary poles [red or purple dots in Fig.~\ref{Fig2}(b)] correspond to bound states with eigenenergies $E_{l}=is_{l}+\omega_{c}$,
\begin{equation}\label{23}
|\varphi_{l}\rangle=(C_{l}|e_{C}\rangle|g_{B}\rangle+B_{l}|g_{C}\rangle|e_{B}\rangle)|vac\rangle+\sum_{k}a^{l}_{k}|g_{C}\rangle|g_{B}\rangle|1^{A}_{k}\rangle.
\end{equation}
The eigenvalues of all other scattering
eigenstates lie in the interval $[\omega_{c}-2J,\omega_{c}+2J]$. 
From the eigenvalue equation $H|\varphi_{l}\rangle=E_{l}|\varphi_{l}\rangle$, the coefficients are determined as
\begin{subequations}\label{24}
	\begin{align}
	C_{l }=\frac{2\lambda^{2}}{N+1}\Sigma_{k}\frac{\sin(\frac{mk\pi}{N+1})\sin(\frac{nk\pi}{N+1})}{E_{l}-\varOmega_{k}},\label{24a}	
	\\B_{l}=E_{l}-\omega_{0}-\frac{2\lambda^{2}}{N+1}\Sigma_{k}\frac{\sin^{2}(\frac{mk\pi}{N+1})}{E_{l}-\varOmega_{k}},\label{24b}
	\\a^{l}_{ k}=\sqrt{\frac{2\lambda^{2}}{N+1}}\frac{C_{l}\sin(\frac{mk\pi}{N+1})+B_{l }\sin(\frac{nk\pi}{N+1})}{E_{l}-\varOmega_{k}},\label{24c}
	\end{align}
\end{subequations}
and
\begin{equation}\label{25}
\frac{\frac{2\lambda^{2}}{N+1}\Sigma_{k}\frac{\sin(\frac{mk\pi}{N+1})\sin(\frac{nk\pi}{N+1})}{E_{l}-\varOmega_{k}}}{E_{l}-\omega_{0}-\frac{2\lambda^{2}}{N+1}\Sigma_{k}\frac{\sin^{2}(\frac{mk\pi}{N+1})}{E_{l}-\varOmega_{k}}}=\frac{E_{l}-\omega_{0}-\frac{2\lambda^{2}}{N+1}\Sigma_{k}\frac{\sin^{2}(\frac{nk\pi}{N+1})}{E_{l}-\varOmega_{k}}}{\frac{2\lambda^{2}}{N+1}\Sigma_{k}\frac{\sin(\frac{mk\pi}{N+1})\sin(\frac{nk\pi}{N+1})}{E_{l}-\varOmega_{k}}},
\end{equation}
 where $C_{l}$, $B_{l}$, and $a^{l}_{k}$ are all real. In the large-$N$ limit,  since $C_{l}=-i\lambda^{2}Q(m,n,s_{l})$ and $B_{l}=is_{l}-\omega+i\lambda^{2}Q(m,m,s_{l})$, Eq.~(\ref{25}) is exactly equivalent to setting the denominators of Eq.~(\ref{11}) to zero. Furthermore, we have
 \begin{equation}\label{26}
 \frac{C_{l}^{2}}{B_{l}^{2}}=\frac{is_{l}-\omega+i\lambda^{2}Q(n,n,s_{l})}{is_{l}-\omega+i\lambda^{2}Q(m,m,s_{l})},
 \end{equation}
 which approaches unity for $|s_{l}|>2J$ as the band-state eigenenergy moves further away from the band edge. In this regime,  the photonic amplitude in position space can be obtained by Fourier transforming Eq.~(\ref{24c}):
 \begin{equation}\label{27}
 \begin{aligned}
 &a^{l}_{j}=\frac{\lambda}{J\sqrt{|\frac{s_{l}}{2J}|^{2}-1}}\Big \{\\
 & C_{l}\Big[sgn[is_{l}]^{|m-j|+1}W(s_{l})^{|m-j|}-sgn[is_{l}]^{m+j+1}W(s_{l})^{m+j}\Big]\\
 &+B_{l}\Big[sgn[is_{l}]^{|n-j|+1}W(s_{l})^{|n-j|}-sgn[is_{l}]^{n+j+1}W(s_{l})^{n+j}\Big]	\Big \},\\
 \end{aligned}
\end{equation}
with $W(s_{l})=|\frac{s_{l}}{2J}|-\sqrt{|\frac{s_{l}}{2J}|^{2}-1}<1$. In deriving this result, we have used the integral identity $\int^{\pi}_{0}\frac{\cos (pk)}{a-\cos (k)}dk=sgn[a]^{p+1}\frac{\pi}{\sqrt{a^{2}-1}}(|a|-\sqrt{a^{2}-1})^{|p|}$, where $a,p\in Z$, $|a|>1$, and $sgn\{\cdot\}$ picks up the sign of ``$\cdot$''.  Ignoring edge effects, the terms proportional to $W(s_{l})^{\eta+j}(\eta=m$ or $n)$ in Eq.~(\ref{27}) are negligible. Consequently, the photonic amplitude is exponentially localized around the QC and QB with a localization length $(\ln [W(s_{l})])^{-1}$ \cite{PhysRevA.96.023831,y5kd-7prs}, and become increasingly extended as $|s_{l}|\rightarrow 2J$. As we show below, the bound states outside the continuum play a key role in energy transfer.

\begin{figure}
		\centering
	\includegraphics[width=0.45\linewidth, height=0.25\textheight]{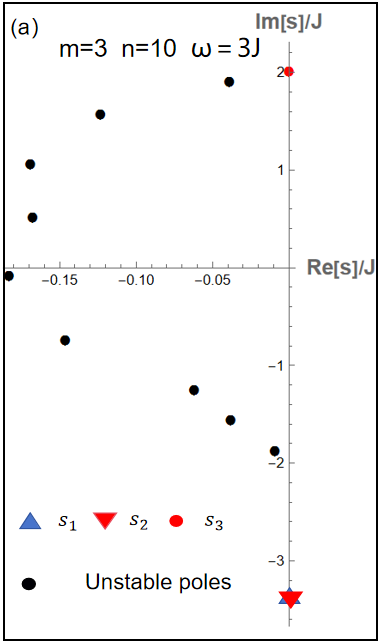}
	\includegraphics[width=0.45\linewidth, height=0.25\textheight]{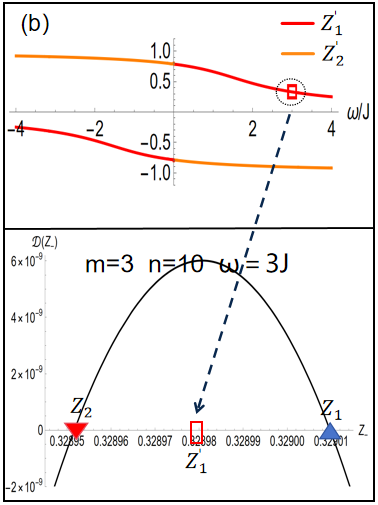}
	\caption{(a) Pole distribution in the complex $s$-plane for $m=3$, $n=10$, $\lambda= J$, and $\omega=3J$. Stable poles $s_1$, $s_2$, and $s_3$ are marked by a blue triangle, a red inverted triangle, and a red dot, respectively; black circles denote unstable poles. Among these, $s_{1}$ and $s_{2}$ located on the same side of the energy band, dominate the charging kinetics. 
	(b) Upper panel: Real roots 
	$Z_{1}^{'}$ (red curve) and $Z_{2}^{'}$ (orange curve) of 
	$h_{0}(\lambda=J,Z_{-}) = 0$ as functions of $\omega$. Lower panel: At $\omega=3J$,  the real roots $Z_{1}$ and $Z_{2}$ of $\mathcal{D}(\lambda=J,Z_{-})=0$ closely approximate $Z^{'}_{1}$.}
	\label{Fig3}
\end{figure}

\begin{figure}
	\centering
	\includegraphics[width=0.43\linewidth, height=0.25\textheight]{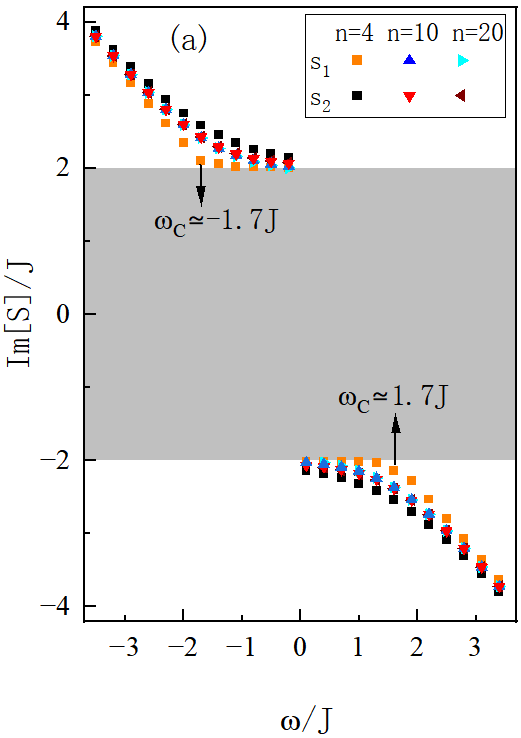}
	\includegraphics[width=0.43\linewidth, height=0.25\textheight]{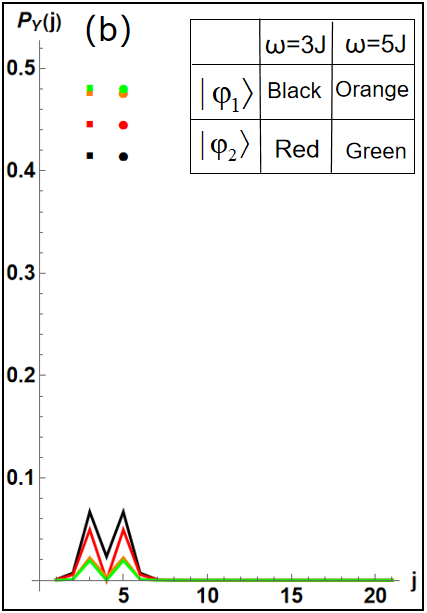}
	\caption{(a) The purely imaginary bound-state singularities $s_{1}$ and $s_{2}$ plotted as functions of the detuning $\omega$ for  $\lambda$= $J$ and $m=3$. As $|\omega|$ increases, $s_{1}$ and $s_{2}$
	approach each other. The shaded region denotes the continuous band. Once $\omega$ crosses the critical value $\omega_C\simeq\pm1.7J$, $s_1$ attaches to the continuum, breaking the two-bound-state interference and suppressing the charging capacity for smaller $m$ and $n$. (b) Normalized spatial profiles $P_{Y}(j)$ of bound eigenstates $|\varphi_1\rangle$ (black/orange) and $|\varphi_2\rangle$ (red/green) under $\omega=3J$ and $5J$. $P_{C}(j)$ (dot), $P_{B}(j)$ (dot) and $P_{W}(j)$ (curve) represent the populations of the QC, QB, and resonator at the $j$-th site, respectively. Other parameters:  $\lambda$= $J$, \(m=3\), \(n=5\).}
	\label{Fig4}
\end{figure}

\section{Bound-State Interference as the Charging Mechanism} 
\subsection{ Interference of Two Out-of-band Bound States}
Our primary goal is to maximize the energy stored in the QB while maintaining a stable charging process. This is achieved at the matched coupling point $\lambda=J$. However, $|\omega|$ must not be too small: as $|\omega|\rightarrow 0$,  $|z(s_{l})|\rightarrow 1$, which suppresses the residue in Eq.~(\ref{22}). For sufficiently large $|\omega|$, after a short transient, the QB and QC settle into stationary, out-of-phase Rabi oscillations with amplitudes approaching
\begin{subequations}\label{28}
	\begin{align}
		B(Jt\gg 1)\simeq\beta_{1}e^{s_{1}t}+\beta_{2}e^{s_{2}t}\label{28a},
		\\
		C(Jt\gg 1)\simeq\gamma_{1}e^{s_{1}t}+\gamma_{2}e^{s_{2}t}\label{28b},
	\end{align}
\end{subequations}
where the purely imaginary poles $s_{1}$ and $s_{2}$, lying on the same side of the cosine band, correspond to the eigenenergies of the two bound states, and $\beta_{1}$, $\beta_{2}$ are their residues. This contrasts with a single emitter coupled to a waveguide, where only one bound state exists on either side of the band \cite{PhysRevA.96.023831}.
In general, three purely imaginary roots exist, but  $s_{3}$ lies so close to the band edge that its residue $\beta_{3}$ is negligible and can be safely omitted.  Fig.~\ref{Fig3}(a) plots all roots  $s_{l}$ of $\mathcal{D}\left[Z_{-}=\frac{is}{2J}- i\sqrt{|1+(\frac{s}{2J})^{2}|}\exp [\frac{Arg(\frac{s}{2J}+i)+Arg(\frac{s}{2J}-i)}{-2i}]\right]=0$ at $\lambda=J$, $\omega=3J$, $m=3$  and $n=10$. The contributions from roots with negative real parts (black dots) decay in the long-time limit. Tracing the origin of $s_{1,2}$ is essential for determining the optimal charging parameters. In fact, the equation $h_{0}(\lambda=J,Z_{-})=0$ from Eq.~(\ref{21a}) has two real roots	$Z_{1}^{'}$ and $Z_{2}^{'}$ with $|Z_{1,2}^{'}|<1$ [Fig.~\ref{Fig3}(b), upper panel]. Upon incorporating higher-order corrections involving $m$ and $n$, the equation $\mathcal{D}(\lambda=J,Z_{-})=0$ supports two real roots $Z_{1}$ and  $Z_{2}$ very close to $Z_{1}^{'}$ [Fig.~\ref{Fig3}(b), lower panel]. Consequently, the corresponding poles are given by $s_{1,2}=-iJ(Z_{1,2}+\frac{1}{Z_{1,2}})$. From Eq.~(\ref{28a}), we have $|B(J\tau)|=\beta_{2}-\beta_{1}$ at $\tau=\frac{(2k-1)\pi}{|s_{1}-s_{2}|}$ ($k=1,2,...$), where $\beta_{2}$ ($\beta_{1}$) is always positive (negative) because the derivatives $\mathcal{D}^{\prime}(Z_{-})|_{Z_{-}=Z_{1}}$ and $\mathcal{D}^{\prime}(Z_{-})|_{Z_{-}=Z_{2}}$ possess opposite signs [see the slope of $\mathcal{D}(Z_{-})$ with respect to $Z_{-}$ in Fig.~\ref{Fig3}(b)]. As shown in Fig.~\ref{Fig4}(a), the frequency splitting $|s_{1}-s_{2}|$ decreases as $|\omega|$ increases, approaching the asymptotic limit $\lim\limits_{\substack{ \omega\to \pm \infty }}s_{1,2}=\mp i\omega$. This indicates that the stored energy in the QB requires a longer time to reach its first peak. Furthermore, for smaller $m$ and $n$, when $\omega$ crosses a critical value $\omega_{C}$, the pole $s_{1}$ with a negligible $\beta_{1}$ may merge into the continuum [see the orange square in  Fig.~\ref{Fig4}(a)]. This destroys the interference between the two bound states, thereby significantly suppressing the charging capability. Fig.~\ref{Fig4}(a) also suggests that increasing the charging distance prolongs the time required for the battery energy to reach its first maximum. Increasing $|\omega|$ will clearly cause the spatial profiles of the two bound states $|\varphi_{1,2}\rangle$ to become more similar, and enhances the weight of QC and QB excited states [cf. Fig.~\ref{Fig4}(b)]. This behavior is quantitatively captured by the normalized population ratio
\begin{equation}\label{29}
\lim\limits_{\substack{	k\to\infty }}\frac{C_{l}^{2}+B_{l}^{2}}{C_{l}^{2}+B_{l}^{2}+\sum_{k}{(a^{l}_{k})}^{2}}\backsimeq\frac{1}{1+\frac{\lambda^{2}}{\pi}	\int^{\pi}_{0}[\frac{\sin (mk)-\sin (nk)}{is_{l}+2J\cos (k)}]^{2}dk},
\end{equation}
where the approximate equality follows from the asymptotic relation $C_{l}\backsimeq-B_{l}$.
\begin{figure}
	\centering
	\includegraphics[width=0.35\linewidth, height=0.2\textheight]{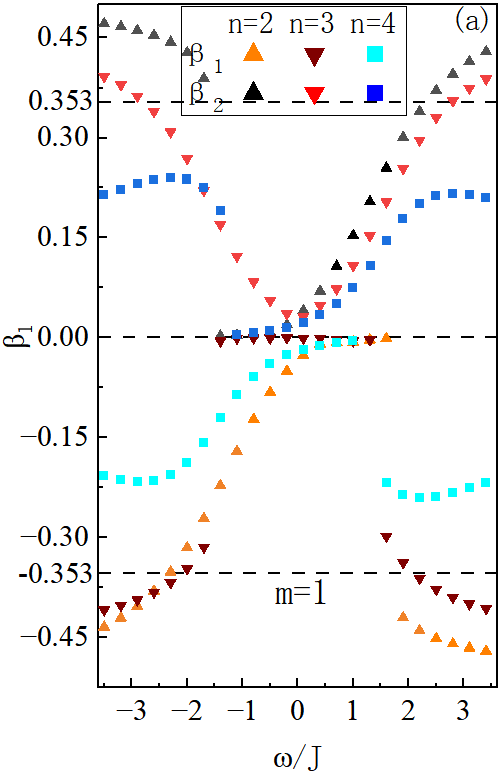}
	\includegraphics[width=0.28\linewidth, height=0.2\textheight]{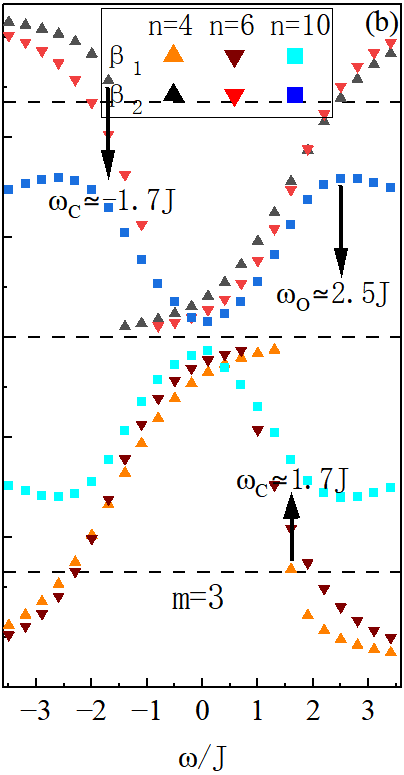}
	\includegraphics[width=0.28\linewidth, height=0.2\textheight]{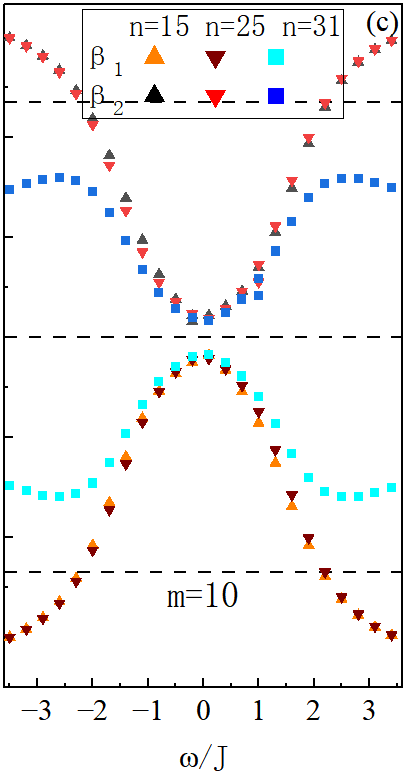}
	\includegraphics[width=0.35\linewidth, height=0.2\textheight]{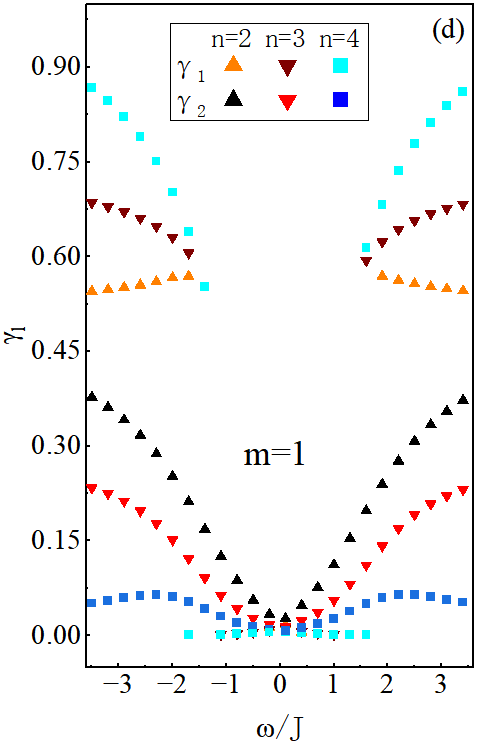}
	\includegraphics[width=0.28\linewidth, height=0.2\textheight]{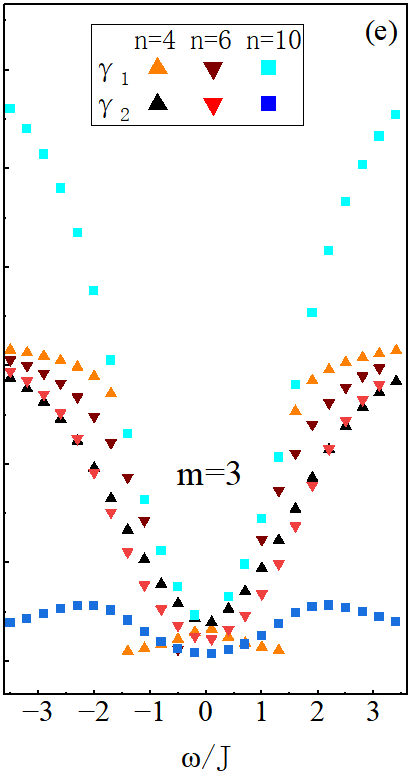}
	\includegraphics[width=0.28\linewidth, height=0.2\textheight]{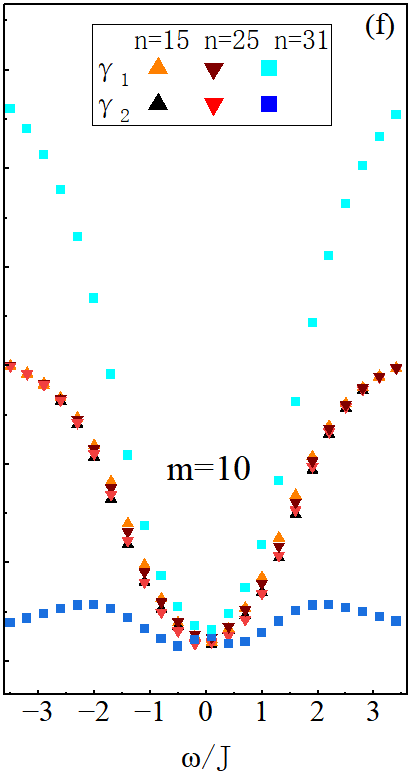}
	\caption{Upper panels (a)–(c): $\beta_{l}$ as functions of the detuning $\omega$ for $m$=1, 3, and 10, respectively. Lower panels (d)–(f): the corresponding $\gamma_{l}$ at the same values of $m$. Color-coded symbols distinguish different site indices $n$, with $\beta_{1,2}$ and $\gamma_{1,2}$ defined in the legends. Panel (b) highlights the critical detuning $|\omega_C|\simeq1.7J$ and the optimal detuning $\omega_O\simeq2.5J$. Effective charging is achieved when $\beta_{2}-\beta_{1}>0.706$.}
	\label{Fig5}
\end{figure}

Fig.~\ref{Fig5}(a,b,c) and Fig.~\ref{Fig5}(d,e,f) respectively illustrate the influence of the detuning $\omega$ on the residues $\beta_{l}$ ($\equiv Res_{B}[s_{l}]$) and $\gamma_{l}$ ($\equiv Res_{C}[s_{l}]$). As direct mappings of the pole $s_{1}$, $\beta_{1}$ and $\gamma_{1}$ 
inherit its characteristic properties, exhibiting an abrupt jump at the critical point $\omega=\omega_{C}$. In general, the relation  $\beta_{1}\simeq-\beta_{2}$ holds universally, with an additional prerequisite $|\omega|>|\omega_{C}|$ when the critical detuning $\omega_{C}$ is well-defined.
Since $\mathcal{D}_{c}(Z_{1})$ and $\mathcal{D}_{c}(Z_{2})$ in Eq.~(\ref{19}) carry opposite signs,  $\gamma_{1}$ and $\gamma_{2}$ always share the same sign.  However, the infinite discontinuity of $\gamma$ in the interval between 
 $Z_{1}$ and $Z_{2}$ leads to a significant discrepancy between  $\gamma_{1}$ and $\gamma_{2}$,  in contrast to the near-antisymmetry of the $\beta$-pair.

\begin{figure}
	\centering
	\includegraphics[width=0.9\linewidth, height=0.18\textheight]{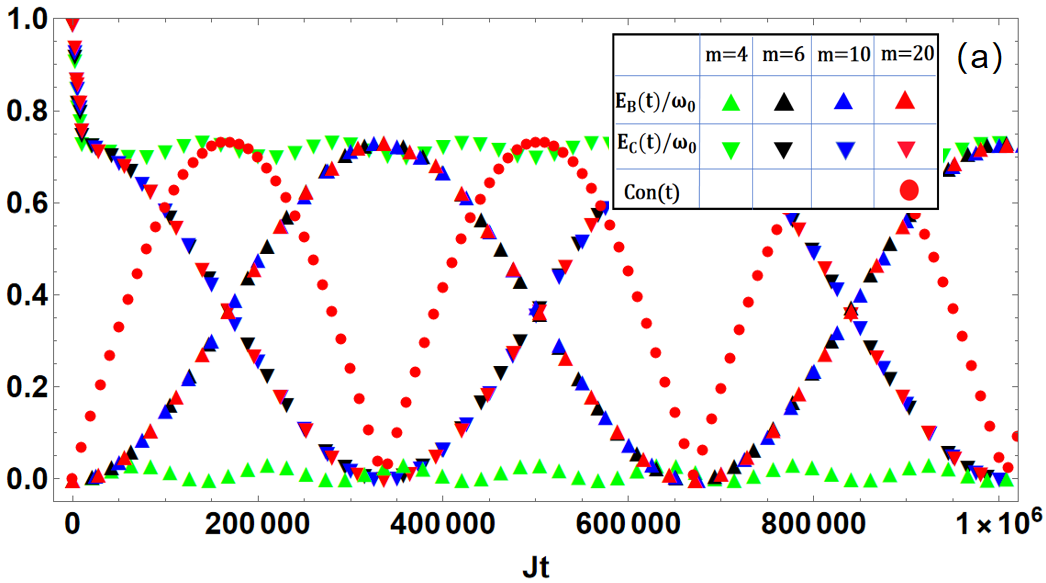}
	\includegraphics[width=0.45\linewidth, height=0.2\textheight]{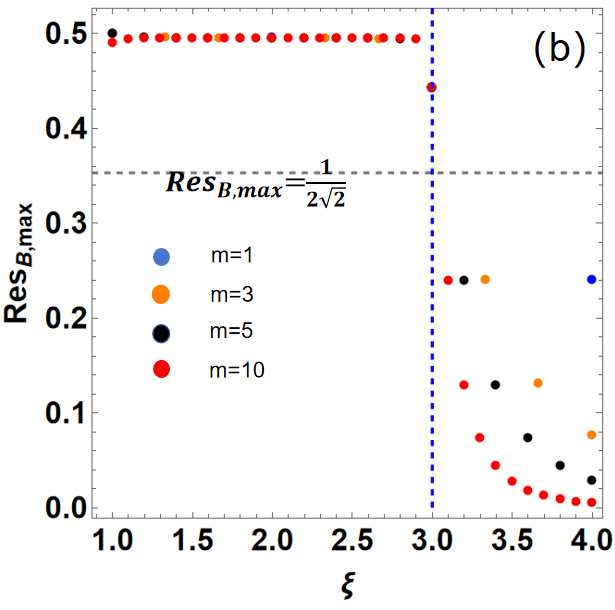}
	\includegraphics[width=0.45\linewidth, height=0.2\textheight]{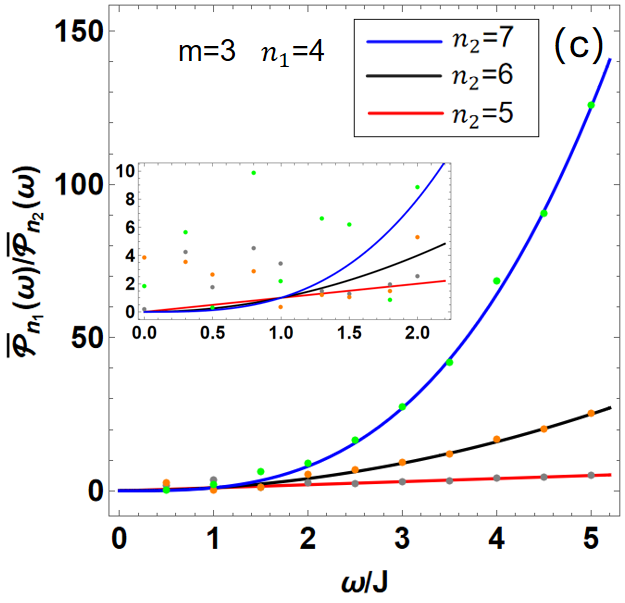}
	\caption{(a) $E_{B}(t)/\omega_{0}$ (triangles), $E_{C}(t)/\omega_{0}$ (inverted triangles), and $Con(t)$ (red dots) versus $Jt$ for $\lambda=J$ and $\omega=3J$. The detuning value is based on the experimental data in Ref.~\cite{PhysRevX.12.031036}; larger detuning leads to greater battery energy storage. Different curves correspond to $m$ = 4 (green), 6 (black), 10 (blue), 20 (red), with the distance $n-m=10$ fixed. (b) Maximum residue $Res_{B,max}$ versus the site ratio $\xi=n/m$ for $m=1$ (blue), 3 (orange), 5 (black), 10 (red). The gray dashed horizontal line indicates the threshold $Res_{B,max}=\frac{1}{2\sqrt{2}}$, and the blue dashed vertical line marks maximum distance of effective charging. (c) Ratio of average powers $\bar{\mathcal{P}}_{n_1}/\bar{\mathcal{P}}_{n_2}$ as a function of the detuning $\omega$ for $m=3$, where
	$n_{1}$ and $n_{2}$ represent different battery positions. The inset provides an enlarged view for $\omega\in[0,2J]$.}
	\label{Fig6}
\end{figure}

\subsection{ Position Effect and Charging Power}

An important feature revealed in Fig.~\ref{Fig5}(a,b,c) is the distinct scaling behavior depending on the relative emitter positions. For $n<3m$, a sufficiently large detuning $|\omega|$ always drives $\beta_{2}-\beta_{1}$ close to unity, indicating that the QB is nearly fully charged. In contrast, when $n>3m$, there exists an optimal detuning $\omega_{O}$ that maximizes $\beta_{2}-\beta_{1}$, yet fails to fully charge the QB. This scaling law is clearly illustrated in Fig.~\ref{Fig6}(a), which shows that charging behavior is nearly identical in the $n<3m$ regime, provided the detuning $\omega$ and the inter-emitter distance $n-m$ remain constant. This stands in stark contrast to the $n>3m$ case, where a significant fraction of the excitation energy remains trapped within the QC [green inverted triangles in Fig.~\ref{Fig6}(a)]. We have also calculated the concurrence \cite{PhysRevLett.80.2245} [red dots in Fig.~\ref{Fig6}(a)] to quantify the entanglement  between the QC and QB, defined as $Con(Jt\gg1)=Max[0,2|C(t)B^{*}(t)|]$. The concurrence vanishes whenever  $C(t)=0$ or $B(t)=0$, occurring at $t=\frac{(2k-1)\pi }{|s_{1}-s_{2}|}$ or $t=\frac{2k\pi }{|s_{1}-s_{2}|}$ ($k=1,2,3...$), respectively. In contrast, the concurrence attains its maximum value $Con(Jt\gg1)_{Max}$ at $t=\frac{\arccos (\phi)+2k\pi}{|s_{1}-s_{2}|}$, where $\phi=-\frac{(\text{$\beta_{1} $} \text{$\gamma_{2} $}+\text{$\beta_{2} $} \text{$\gamma_{1} $}) (\text{$\beta_{1} $} \text{$\gamma_{1} $}+\text{$\beta_{2} $} \text{$\gamma_{2} $})}{4 \text{$\beta_{1} $} \text{$\beta_{2} $} \text{$\gamma_{1} $} \text{$\gamma_{2} $}}$. At these instants, the instantaneous charging and discharging powers are maximized, and the QC and QB share equal energy.
	
The dependence of the residues $\beta_{1,2}$ on the charging distance is governed primarily by the numerator $i\mathcal{D}_{B}(Z_{1,2})$ in Eq.~(\ref{22}), which diminishes as the separation increases. For the QB to deliver external work, a positive ergotropy $\mathcal{E}(\tau_{1})>0$ is required, which in turn demands $\beta_{2}\backsimeq-\beta_{1}>\frac{1}{2\sqrt{2}}$—a condition that cannot be satisfied by tuning the detuning once $n>3m$. To confirm this, we set $\mathcal{D}(Z_{-})=0$ in Eq.~(\ref{19}) to express the detuning as $\omega=f(Z_{-})$. Substituting this expression into Eq.~(\ref{22}) and scanning $Z_{-}$ over $(-1,1)$---corresponding to the poles $s_{1,2}$ moving from the band edge deep into the off-band region, i.e., 
$|Z_{-}|$ decreasing from $1$ to $0$---we determine the maximum residue $Res_{B,max}$ for fixed $m$ and $n$. As shown in Fig.~\ref{Fig6}(b), $Res_{B,max}>\frac{1}{2\sqrt{2}}$ holds generically for any $m$ and $n$ satisfying $n<3m$, whereas $Res_{B,max}$ never exceeds  $\frac{1}{2\sqrt{2}}$ when $n>3m$. Although the QB can in principle be charged to near full capacity for $n<3m$,
the time required for the QB to reach its first local maximum will grow exponentially with the separation. In fact, under large detuning $|\omega|\gg J$, Fig.~\ref{Fig6}(a) shows that the poles $s_{1,2}$ become insensitive to the QC position $m$ once the inter-emitter separation $n-m$ is fixed. In this regime, we can derive an empirical scaling relation:
\begin{equation}\label{30}
	\frac{ \overline{\mathcal{P}}_{n_{1}}(\omega)}{ \overline{\mathcal{P}}_{n_{2}}(\omega)}\simeq\frac{\tau_{1}(n_{2},\omega)}{\tau_{1}(n_{1},\omega)}\simeq(\omega/J)^{n_{2}-n_{1}},
\end{equation}
valid for $m<n_{1}<n_{2}<3m$. Eq.~(\ref{30}) is confirmed by extensive numerical simulations [Fig.~\ref{Fig6}(c)], with deviations arising predominantly from the first approximate equality at small $|\omega|$ [inset in Fig.~\ref{Fig6}(c)].

\begin{figure}
	\centering
	\includegraphics[width=0.9
	\linewidth]{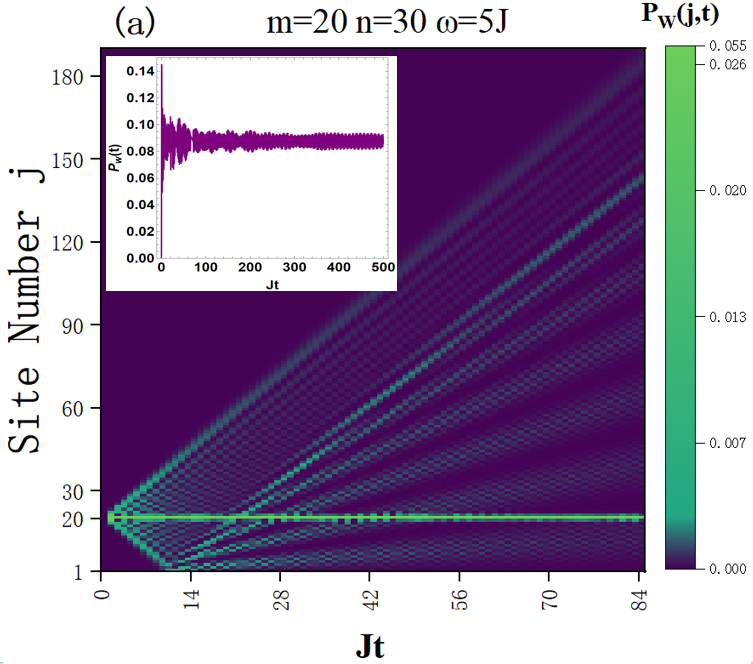}
	\includegraphics[width=0.9
	\linewidth]{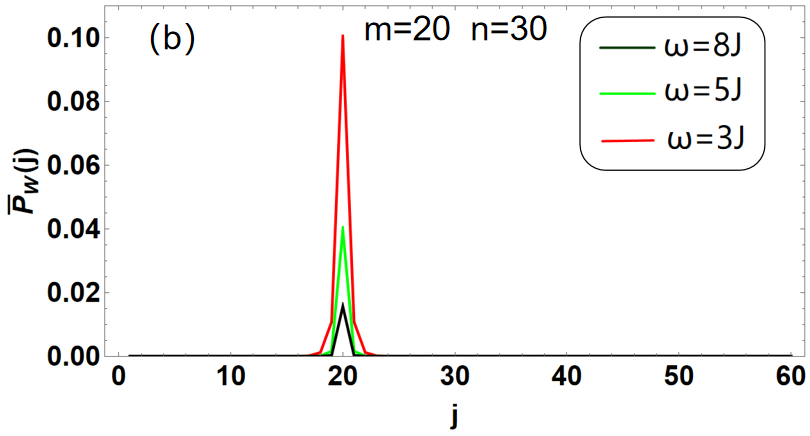}
	\caption{(a) 
		Spatiotemporal evolution of the photon density $P_{W}(j,t)$ for $m$=20, $n$=30, and detuning $\omega$=$5J$. The inset plots the total photon density, summed over all sites $j$, as a function of time.
		(b) Long-time-averaged photon density $\overline{P}_{W}(j)$ versus site index $j$ for $m$=20, $n$=30, with detuning $\omega$ = $3J$ (red), $5J$ (green), $8J$ (black). Photons are strongly localized near the lattice site \(j=m\), and the peak amplitude of the averaged photon density gradually decreases with increasing $\omega$.}
	\label{Fig7}
\end{figure}

\subsection{Photon Probability Distribution}

Due to the contribution from the Hankel-path integrals along the contours $C_{1,2,3,4}$ and the unstable poles located far from the imaginary axis in Eq.~(\ref{17}),  a small fraction of the initial excitation is released into the waveguide within an extremely short time interval [inset of Fig.~\ref{Fig7}(a)]. The sustained yet minute fluctuations in the waveguide photon probability 
$P_{W}(t)\equiv1-|C(t)|^{2}-|B(t)|^{2}$ arise from the interplay between the third stable pole $s_{3}$ and other unstable poles lying near the imaginary axis. The spatiotemporal evolution of the photon probability density, defined as $P_{W}(j,t)(\equiv\big|\Sigma_{k=1}^{N}{\sqrt\frac{2}{N+1}}\sin (\frac{jk\pi}{N+1})a_{k}(t)\big|^{2})$, is plotted in Fig.~\ref{Fig7}(a). While the majority of the emitted radiation remains localized around the QC, a small fraction propagates symmetrically outward from the QC at a maximum group velocity of $\frac{|j-m|}{t}\backsimeq2J$ \cite{PhysRevA.89.053826} and is reflected at the near end. The presence of the QB has negligible impact on the diffusion dynamics of the radiation field. As the light propagates along the waveguide,  $P_{W}(j,t)$ exhibits only minor fluctuations around its long-time average, which can be evaluated exactly in the large-$N$ limit via Eq.~(\ref{7}) and Eq.~(\ref{28}):
\begin{equation}\label{31}
\begin{aligned}
&\overline{P}_{W}(j)=\lim\limits_{\substack{\tau\to \infty }}\frac{1}{\tau}\int_{0}^{\tau}P_{W}(j,t) dt\\
&=-\lambda^{2}[(\beta_{1}+\beta_{2})Q(n,j,s_{1})+(\gamma_{1}+\gamma_{2})Q(m,j,s_{1})]\\
&[(\beta_{1}+\beta_{2})Q(n,j,s_{2})+(\gamma_{1}+\gamma_{2})Q(m,j,s_{2})].\\
\end{aligned}
\end{equation}
Since $\beta_{1}+\beta_{2}\simeq0$, the expression simplifies to $\overline{P}_{W}(j)\backsimeq-\lambda^{2}(\gamma_{1}+\gamma_{2})^{2}Q(m,j,s_{1})Q(m,j,s_{2})$. Given $|z_{-}(s_{1,2})|<1$, the first term in the numerator of $Q_{-}$ in Eq.~(\ref{15}) can generally be neglected, while the second term accurately captures the localization behavior around the QC. Fig.~\ref{Fig7}(b) displays the time-averaged photon probability density for various detuning values $\omega$. Because 
$\gamma_{1}+\gamma_{2}$ is largely  insensitive to $\omega$, it is evident that reducing $\omega$ amplifies the magnitude of $|z_{-}(s_{1,2})|$, thereby enhancing the population trapping.

\section{Origin of Out-of-Band Bound States and  Parameter Optimization}
\begin{figure}
	\centering
	\includegraphics[width=0.9\linewidth]{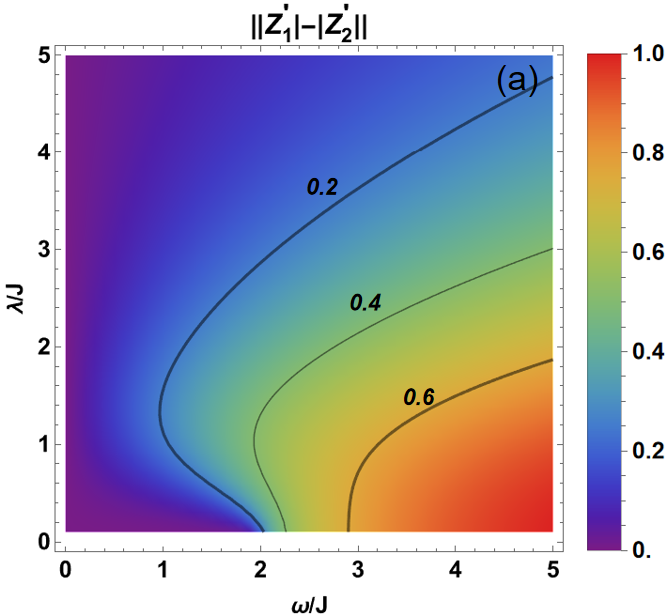}
	\includegraphics[width=0.45\linewidth]{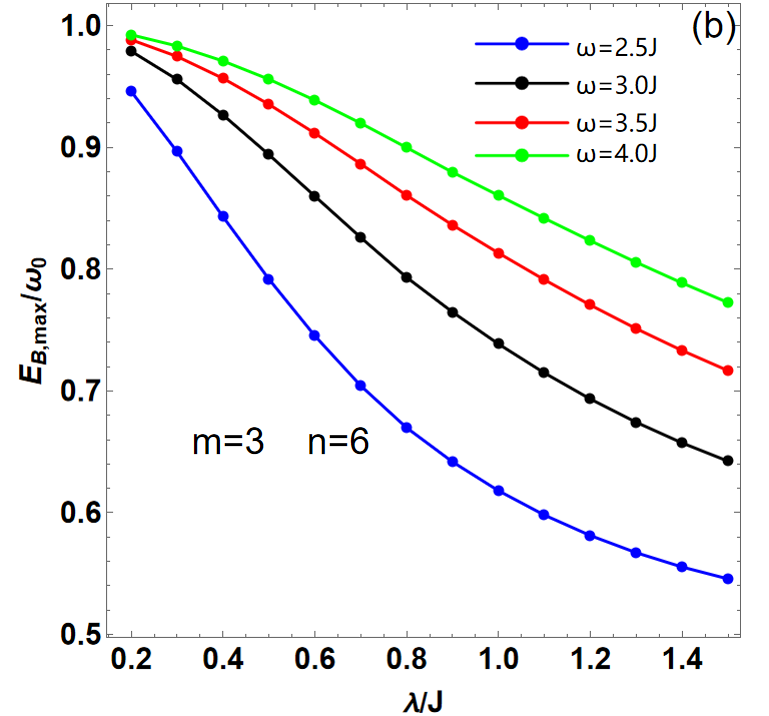}
	\includegraphics[width=0.45\linewidth]{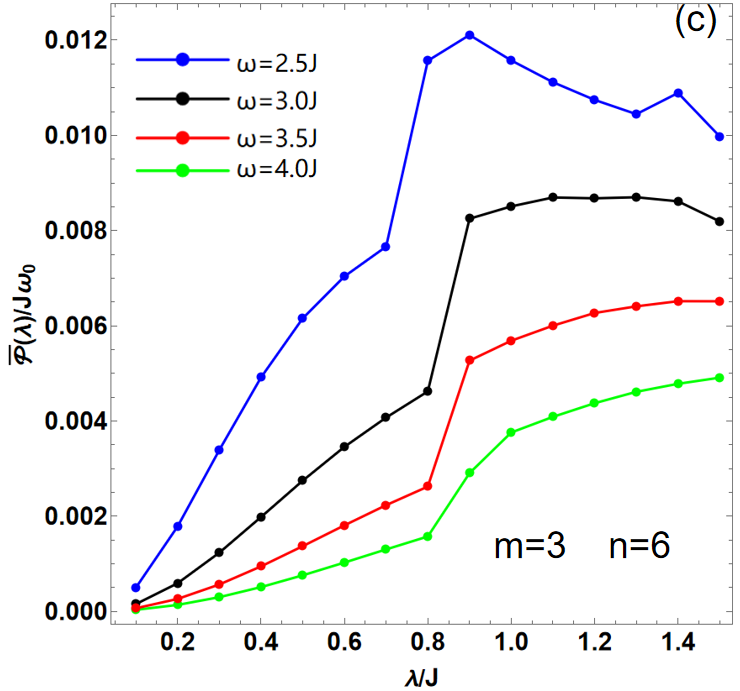}
	\caption{(a) $||Z^{'}_{1}|-|Z^{'}_{2}||$ as a function of $\lambda$ and  $\omega$. Stable charging is dominated by two bound states only in regions where $||Z^{'}_{1}|-|Z^{'}_{2}||$ is large. (b) and (c)  show $E_{B,max}$ and $\mathcal{\overline{P}(\lambda)}$ versus coupling strength $\lambda$ for different detunings $\omega$ in the large-$||Z^{'}_{1}|-|Z^{'}_{2}||$ regime. $\omega$=$2.5J$ (blue), $3.0J$ (black), $3.5J$ (red), $4.0J$ (green). $m=3$ and $n=6$ are fixed.}
	\label{Fig8}
\end{figure}
 
In the previous section, we identified that the two bound states on the same side of the energy band play a pivotal role in enabling the stable charging protocol. Although this mechanism does not strictly require the coupling-matched condition $\lambda=J$, the system parameters remain subject to certain constraints. For any out-of-band bound state, the corresponding real root $Z_{l}$ of $\mathcal{D}(Z_{-})=0$ must satisfy  $|Z_{l}|<1$, as illustrated in  Fig.~\ref{Fig2}. In fact, all such real roots $Z_{l}$ originate from the two real solutions of $h_{0}(Z_{-}) = 0$ defined in Eq.~(\ref{21a}), within the interval $(-1, 1)$. Following the notation of the previous section, we denote these two roots---one positive and one negative---as $Z_{1}^{'}$ and $Z_{2}^{'}$ for $\omega>0$. If $\omega$ changes sign to $-\omega$, the roots transform as $Z_{1,2}^{'}\rightarrow-Z_{2,1}^{'}$. When the correction terms involving $m$ and $n$ are included,  $\mathcal{D}(Z_{-})=0$ typically yields four real roots, with $Z_{1,2}$ and $Z_{3,4}$ located near  $Z^{'}_{1}$ and $Z^{'}_{2}$, respectively. The condition $|Z_{l}|<1$ implies that the number of bound states can be two, three, or four. In the strongly coupled regime $\lambda\gg J$, all four bound states contribute significantly, and the temporal requirements for achieving resonance among them become extremely stringent. 

The optimal stabilization mechanism is one in which only two bound states dominate the dynamics, and the normalized stored energy $E_{B}(t_{max})/\omega_{0}$
approaches unity as closely as possible when these two states resonate. A viable parameter selection strategy is to set $|Z^{'}_{1}|$ close to $0$ and  $|Z^{'}_{2}|$ close to $1$. With such a setting, any bound state associated with $Z_{3}$ or $Z_{4}$---if it exists---contributes negligibly to the dynamics. The two bound states on the same side of the energy band, corresponding to $Z_{1,2}$, achieve stable charging via coherent resonance. Moreover, since $|Z_{1,2}|\ll 1$, the eigenenergies of these bound states remain largely insensitive to variations in the positions of the QC and QB. Consequently, the approximate parameter regime can be determined using the criterion $||Z^{'}_{1}|-|Z^{'}_{2}||\gtrsim 0.5$ , as illustrated in Fig.~\ref{Fig8}(a).  However, Fig.~\ref{Fig8}(b) and Fig.~\ref{Fig8}(c) demonstrate that within this regime, it is not possible to simultaneously maximize both the stored energy $E_{B,max}$ and the average charging power $\overline{\mathcal{P}}$.

\section{Ineffectiveness of In-Band Bound States}

 For a specific detuning $\omega=2J \cos (\frac{\pi l}{n-m})$ (where $l$ is an integer satisfying $-2J<\omega<2J$ and $\omega\neq0$), the pole $s_{l}=-2iJ \cos (\frac{\pi l}{n-m})$ [e.g., the purple dot in the Fig.~\ref{Fig2}(b)] signals a  bound state in the continuum. The ratio $C_{l}^{2}/B_{l}^{2}$ in  Eq.~(\ref{26}) becomes highly sensitive to system parameters, because under this condition, the complex $z_{-}(s_{l})\equiv e^{\frac{i\pi l}{n-m}}$ lies on the unit circle. Eq.~(\ref{22}) yields the corresponding residues:

\begin{equation}\label{32}
\begin{aligned}
Res_{C}[s=-2iJ \cos (\frac{\pi l}{n-m})]=\frac{2 \sin^{2}(\frac{\pi l}{n-m})J^{2}}{(n-m)\lambda^{2}+4\sin^{2}(\frac{\pi l}{n-m})J^{2} },		
\\		
Res_{B}[s=-2iJ \cos (\frac{\pi l}{n-m})]=\frac{(-1)^{l+1} 2 \sin^{2}(\frac{\pi l}{n-m})J^{2}}{(n-m)\lambda^{2}+4\sin^{2}(\frac{\pi l}{n-m})J^{2} }.
\end{aligned}
\end{equation}
The residues at zero detuning ($\omega=0$) depend on the parity of $m$ and $n$.

For odd $m$ and $n$:
\begin{equation}\label{33}
Res_{C}[s=0]=\frac{1}{2+\frac{(n-m)\lambda^{2}}{2J^{2}}}, Res_{B}[s=0]=
\frac{-i^{n-m}}{2+\frac{(n-m)\lambda^{2}}{2J^{2}}}.
\end{equation}
\\For even $m$ and $n$:
\begin{equation}\label{34}
\begin{aligned}
Res_{C}[s=0] &=\frac{2 (J^{4}+\frac{n}{2}J^{2}\lambda^{2})}{2J^{4}+(m+n)J^{2}\lambda^{2}+\frac{m(n-m)}{2}\lambda^{4}},\\
Res_{B}[s=0]&=\frac{-2(i)^{n+m}J^{2}\lambda^{2}}{\frac{4}{m}J^{4}+\frac{2(m+n)}{m}J^{2}\lambda^{2}+(n-m)\lambda^{4}}.
\end{aligned}
\end{equation}
\\For even $m$ and odd $n$:
\begin{equation}\label{35}
Res_{C}[s=0] =\frac{1}{1+\frac{m\lambda^{2} }{2J^{2}}},Res_{B}[s=0]=0.
\end{equation}
\\For odd $m$ and even $n$, the pole at $s=0$ is absent. In general, the photonic population in these states is predominantly localized between the QC and the QB.  If a specific $k^{\ast}$  meets $\cos (\frac{\pi k^{\ast}}{N+1}) = \cos(\frac{\pi l}{n-m})$, we obtain $C_{l}=-B_{l}$, $a^{l}_{k^{\ast}}=\sqrt{\frac{2\lambda^{2}}{N+1}}B_{l}[-\sin(\frac{mk^{\ast}\pi}{N+1})+\sin(\frac{nk^{\ast}\pi}{N+1})]$, and $a^{l}_{k\neq k^{\ast}}=0$. This indicates that the state is precisely a Bloch wave propagating along the CRW.

\begin{figure}
	\centering
	\includegraphics[width=0.86\linewidth]{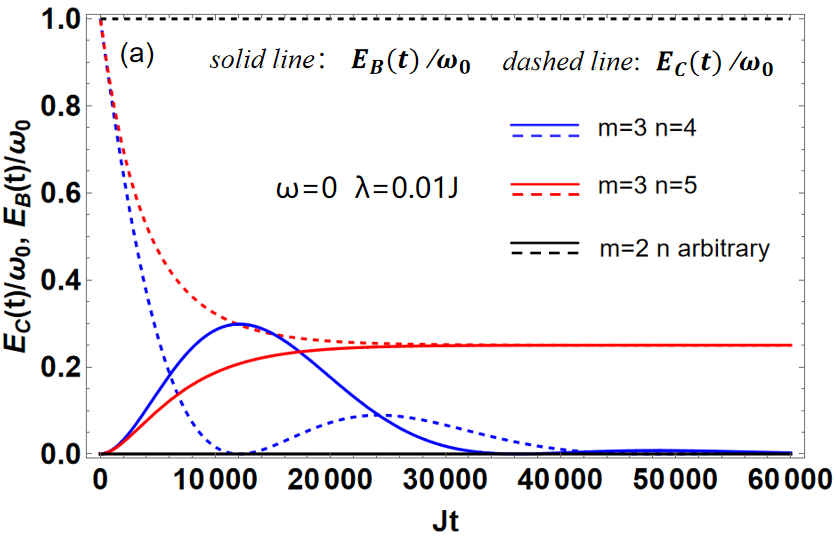}
	\includegraphics[width=0.85\linewidth]{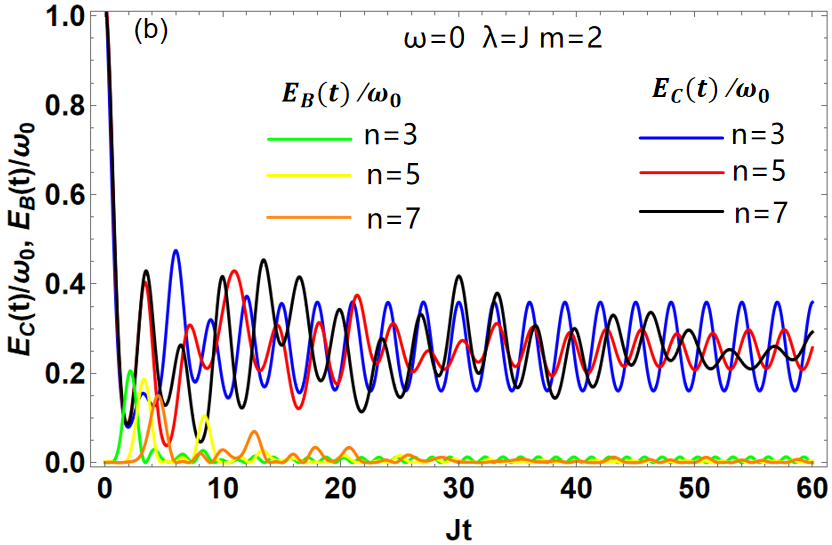}
	\caption{Dynamics of $E_{B}(t)$ and $E_{C}(t)$ at $\omega=0$.  The site indices $m$ and $n$ are specified in the figure legend. (a) In the ultra-weak coupling limit, $\lim _{\lambda\rightarrow 0}E_{C}(t,n)=1$ and $\lim _{\lambda\rightarrow 0}E_{B}(t,n)=0$   for even $m$. (b) In the coupling-matched regime, $E_{C}(t)/\omega_{0}$ oscillates around $Res_{C}[s=0]$.}
	\label{Fig9}
\end{figure}

Note that the residues in Eqs.~(\ref{32}) and (\ref{33}) are determined solely by the QC--QB separation, while those in Eqs.~(\ref{34}) and (\ref{35}) are influenced by the presence of the boundary.
Fig.~\ref{Fig9}(a) illustrates that minor adjustments to $m$ and $n$ 
 result in fundamentally distinct dynamical behaviors governed by
in-band bound states at $\omega=0$.  In the single-emitter scenario, Markovian dynamics typically dominates when both $|\omega|$ and $\lambda$ are much smaller than $J$ \cite{PhysRevA.89.053826}. Here, the configuration $m=3,n=5$ produces a subradiant state, whereas $m=3,n=4$ leads to a superradiant state—both consistent with the PBC case \cite{10.1088/1674-1056/ae3b32}. In contrast, when $m=2$, the energy is trapped in the QC. This feature is independent of the QB's position in the ultra-weak coupling limit. For sufficiently large $m$ (which effectively recovers the infinite-waveguide limit), $Res_{C}[s=0]$ in Eqs.~(\ref{34}) and (\ref{35}) vanishes, implying that the energy fully dissipates into the waveguide. Increasing $\lambda$ induces additional out-of-band bound states, causing  $E_{B,C}(t)/\omega_{0}$ to oscillate around a smaller value of  $|Res_{B,C}[s=0]|^{2}$, as shown in Fig.~\ref{Fig9}(b). However, given that 
Eqs.~(\ref{32}-\ref{35}) bound $|Res_{B}|<0.5$ and  that out-of-band and in-band bound states 
prevail in distinct parameter regimes, the latter cannot be employed for charging optimization.

\section{CONCLUSION}
In this work, we investigate energy transfer between two emitters mediated by a semi-infinite CRW and identify the parameter regime for optimal charging, corresponding to the high-value region of $||Z^{'}_{1}|-|Z^{'}_{2}||$ in Fig.~\ref{Fig8}(a). In this regime, stable long-range periodic charging is achieved through the interference of two bound states on the same side of the energy band, yet the stored energy and the average charging power of the QB cannot be simultaneously optimized. Notably, the two bound states exhibit comparable photon localization at the QC and QB sites on the waveguide, in sharp contrast to the real-time excitation density, which remains predominantly localized near the QC.

While the QB can in principle be charged over arbitrary distances via the CRW with positive extractable work, the breaking of translational symmetry constrains the relative positioning of QC and QB. For instance, taking $\lambda=J$,  when $n< 3m$, full charging of the QB is achievable by increasing the detuning magnitude, although a larger detuning delays the first local maximum. Conversely, for $n>3m$, extracting positive work from the QB becomes impossible. Moreover, in the regime $n<3m$ with large detuning, the average charging power decays exponentially with increasing charging distance. These scaling behaviors are not limited to $\lambda=J$, but extend to the entire parameter regime where the dynamics are dominated by the two bound states on the same band edge. In the strong-coupling regime ($\lambda\gg J$), the time at which multimode resonance occurs becomes prohibitively sensitive to system parameters, rendering stable charging impractical. For weak coupling strengths, the dynamics exhibit
 a pronounced dependence on the parity of the QC and QB site indices at $\omega=0$. This positional effect, together with the site-dependent constraints mentioned above, stems from end-point reflections under OBCs. 

Coherent dynamics dominated by bound states can be faithfully reproduced in a finite-length waveguide, once its length greatly exceeds the spatial extent of the photon cloud \cite{PhysRevApplied.20.024058,Mirhosseini2018}. Our results therefore have direct experimental implications, particularly for the integration of QB in SQCs. In the experiment of Ref.\cite{PhysRevX.12.031036}, each resonator consists of an array of ten JJs shunted by a capacitor, with transition frequency $\omega_{c}/2 \pi \approx 5.593$ GHz and nearest-neighbor coupling $J/2\pi \approx 249$ MHz, while the QC and QB can be emulated by flux-tunable superconducting transmons operating at a maximum frequency $\omega_{0}/2\pi \approx 6.33$ GHz and coupled with strength  $\lambda/2 \pi \approx 311$ MHz. These parameters correspond to a detuning $\omega=\omega_{0}-\omega_{c} \approx 3J$ in our reduced units, with one unit of time $1Jt\approx 0.64~ns$; the system can thus support an effective charging distance of $n-m=10$ [Ref. Fig.~\ref{Fig6} (a)], bringing the QB to its first energy maximum at approximately $280~\mu s$—a timescale on the order of the coherence time of state-of-the-art transmon platforms \cite{Tuokkola2025,Zhang2022,Acharya2025,bland2025millisecond}.  Upon full charging, the waveguide returns to the vacuum state. This allows energy to be relayed to more distant qubits in multi-node networks on experimentally accessible timescales by sequentially modulating the local couplings $\lambda$ via a relay protocol. These insights provide a useful perspective for the design of quantum energy transmission networks.

 \section*{Acknowledgments} 
This work was supported by the National Science Foundation of China (Grant No. 11664021, 61565008, 11365013), by the Opening Foundation of State Key Laboratory of Surface Physics (Grant No.KF2017\underline{ }06),  and by Yunnan Ten Thousand Talents Plan Young and Elite Talents Project (Grant Number:YNWR-QNBJ-2018-121).
\appendix
\renewcommand{\appendixname}{APPENDIX: HANKEL PATH INTEGRALS ALONG THE BRANCH CUTS}
\renewcommand{\thesection}{}
\makeatletter
\def\@seccntformat#1{}
\makeatother 
\setcounter{equation}{0}
\renewcommand{\theequation}{A\arabic{equation}}
\section{}
By setting $s=x\pm 2iJ$, the integrals along the four branch cuts $C_{1,2,3,4}$ are explicitly given by
\begin{widetext}\label{A1}
	\begin{subequations}
		\begin{align}			
			\lim\limits_{\substack{ r\to \ 0\\ }}\int_{C_{1}}B(s)e^{st}ds=\int^{0}_{-\infty}\frac{\frac{i\lambda^{2}(W_{1}^{1+m+n}-W_{1}^{1-m+n})}{J(W_{1}^{2}-1)}e^{(x+2iJ)t}dx}{\frac{-\lambda^{4}(W_{1}^{1+m+n}-W_{1}^{1-m+n})^{2}}{J^{2}(W_{1}^{2}-1)^{2}}-[\frac{i\lambda^{2}(W_{1}^{1+2n}-W_{1})}{J(W_{1}^{2}-1)}+x+2iJ+i\omega][\frac{i\lambda^{2}(W_{1}^{1+2m}-W_{1})}{J(W_{1}^{2}-1)}+x+2iJ+i\omega]},\\
			\lim\limits_{\substack{ r\to \ 0\\ }}\int_{C_{1}}C(s)e^{st}ds=\int^{0}_{-\infty}\frac{[\frac{-i\lambda^{2}(W_{1}^{1+2n}-W_{1})}{J(W_{1}^{2}-1)}+x+2iJ+i\omega]e^{(x+2iJ)t}dx}{\frac{-\lambda^{4}(W_{1}^{1+m+n}-W_{1}^{1-m+n})^{2}}{J^{2}(W_{1}^{2}-1)^{2}}-[\frac{i\lambda^{2}(W_{1}^{1+2n}-W_{1})}{J(W_{1}^{2}-1)}+x+2iJ+i\omega][\frac{i\lambda^{2}(W_{1}^{1+2m}-W_{1})}{J(W_{1}^{2}-1)}+x+2iJ+i\omega]},\\
			\lim\limits_{\substack{ r\to \ 0\\ }}\int_{C_{2}}B(s)e^{st}ds=\int^{0}_{-\infty}\frac{\frac{-i\lambda^{2}(W_{2}^{1+m+n}-W_{2}^{1-m+n})}{J(W_{2}^{2}-1)}e^{(x+2iJ)t}dx}{\frac{-\lambda^{4}(W_{2}^{1+m+n}-W_{2}^{1-m+n})^{2}}{J^{2}(W_{2}^{2}-1)^{2}}-[\frac{i\lambda^{2}(W_{2}^{1+2n}-W_{2})}{J(W_{2}^{2}-1)}+x+2iJ+i\omega][\frac{i\lambda^{2}(W_{2}^{1+2m}-W_{2})}{J(W_{2}^{2}-1)}+x+2iJ+i\omega]},\\
			\lim\limits_{\substack{ r\to \ 0\\ }}\int_{C_{2}}C(s)e^{st}ds=\int^{0}_{-\infty}\frac{-[\frac{-i\lambda^{2}(W_{2}^{1+2n}-W_{2})}{J(W_{2}^{2}-1)}+x+2iJ+i\omega]e^{(x+2iJ)t}dx}{\frac{-\lambda^{4}(W_{2}^{1+m+n}-W_{2}^{1-m+n})^{2}}{J^{2}(W_{2}^{2}-1)^{2}}-[\frac{i\lambda^{2}(W_{2}^{1+2n}-W_{2})}{J(W_{2}^{2}-1)}+x+2iJ+i\omega][\frac{i\lambda^{2}(W_{2}^{1+2m}-W_{2})}{J(W_{2}^{2}-1)}+x+2iJ+i\omega]},\\
			\lim\limits_{\substack{ r^{\prime}\to \ 0\\ }}\int_{C_{3}}B(s)e^{st}ds=\int^{0}_{-\infty}\frac{\frac{i\lambda^{2}(W_{3}^{1+m+n}-W_{3}^{1-m+n})}{J(W_{3}^{2}-1)}e^{(x-2iJ)t}dx}{\frac{-\lambda^{4}(W_{3}^{1+m+n}-W_{3}^{1-m+n})^{2}}{J^{2}(W_{3}^{2}-1)^{2}}-[\frac{i\lambda^{2}(W_{3}^{1+2n}-W_{3})}{J(W_{3}^{2}-1)}+x-2iJ+i\omega][\frac{i\lambda^{2}(W_{3}^{1+2m}-W_{3})}{J(W_{3}^{2}-1)}+x-2iJ+i\omega]},\\
			\lim\limits_{\substack{r^{\prime} \to \ 0\\ }}\int_{C_{3}}C(s)e^{st}ds=\int^{0}_{-\infty}\frac{[\frac{-i\lambda^{2}(W_{3}^{1+2n}-W_{3})}{J(W_{3}^{2}-1)}+x-2iJ+i\omega]e^{(x-2iJ)t}dx}{\frac{-\lambda^{4}(W_{3}^{1+m+n}-W_{3}^{1-m+n})^{2}}{J^{2}(W_{3}^{2}-1)^{2}}-[\frac{i\lambda^{2}(W_{3}^{1+2n}-W_{3})}{J(W_{3}^{2}-1)}+x-2iJ+i\omega][\frac{i\lambda^{2}(W_{3}^{1+2m}-W_{3})}{J(W_{3}^{2}-1)}+x-2iJ+i\omega]},\\
			\lim\limits_{\substack{ r^{\prime}\to \ 0\\ }}\int_{C_{4}}B(s)e^{st}ds=\int^{0}_{-\infty}\frac{\frac{-i\lambda^{2}(W_{4}^{1+m+n}-W_{4}^{1-m+n})}{J(W_{4}^{2}-1)}e^{(x-2iJ)t}dx}{\frac{-\lambda^{4}(W_{4}^{1+m+n}-W_{4}^{1-m+n})^{2}}{J^{2}(W_{4}^{2}-1)^{2}}-[\frac{i\lambda^{2}(W_{4}^{1+2n}-W_{4})}{J(W_{4}^{2}-1)}+x-2iJ+i\omega][\frac{i\lambda^{2}(W_{4}^{1+2m}-W_{4})}{J(W_{4}^{2}-1)}+x-2iJ+i\omega]},\\
			\lim\limits_{\substack{r^{\prime} \to \ 0\\ }}\int_{C_{4}}C(s)e^{st}ds=\int^{0}_{-\infty}\frac{-[\frac{-i\lambda^{2}(W_{4}^{1+2n}-W_{4})}{J(W_{4}^{2}-1)}+x-2iJ+i\omega]e^{(x-2iJ)t}dx}{\frac{-\lambda^{4}(W_{4}^{1+m+n}-W_{4}^{1-m+n})^{2}}{J^{2}(W_{4}^{2}-1)^{2}}-[\frac{i\lambda^{2}(W_{4}^{1+2n}-W_{4})}{J(W_{4}^{2}-1)}+x-2iJ+i\omega][\frac{i\lambda^{2}(W_{4}^{1+2m}-W_{4})}{J(W_{4}^{2}-1)}+x-2iJ+i\omega]},
		\end{align}
	\end{subequations}
where  $W_{1}=[\frac{xi}{2J}-1+(\frac{x^{2}}{J^{2}}+\frac{x^{4}}{16J^{4}})^{\frac{1}{4}}e^{\frac{i\theta(x)}{2}}], W_{2}=[\frac{xi}{2J}-1-(\frac{x^{2}}{J^{2}}+\frac{x^{4}}{16J^{4}})^{\frac{1}{4}}e^{\frac{i\theta(x)}{2}}], W_{3}=[\frac{xi}{2J}+1+(\frac{x^{2}}{J^{2}}+\frac{x^{4}}{16J^{4}})^{\frac{1}{4}}e^{\frac{-i\theta(x)}{2}}], W_{4}=[\frac{xi}{2J}+1-(\frac{x^{2}}{J^{2}}+\frac{x^{4}}{16J^{4}})^{\frac{1}{4}}e^{\frac{-i\theta(x)}{2}}]$ and $\theta(x)=Arg(\frac{x}{2J}+2i)$.
The exponentially decaying factor $e^{xt}$ renders the four integrals negligible for $Jt \gtrsim 20$ under the coupling matching and large detuning.
\end{widetext}

\bibliographystyle{apsrev4-2}
\bibliography{Maintextreference}

\end{document}